\documentclass[11pt]{cernrep}
\usepackage{graphicx}
\usepackage{epsf}
\usepackage{epsfig}
\def\lapproxeq{\lower .7ex\hbox{$\;\stackrel{\textstyle <}{\sim}\;$}}
\def\gapproxeq{\lower .7ex\hbox{$\;\stackrel{\textstyle >}{\sim}\;$}}

\def\be{\begin{equation}}
\def\ee{\end{equation}}
\def\bea{\begin{eqnarray}}
\def\eea{\end{eqnarray}}

\begin{document}
\begin{titlepage}
\begin{flushright}
IPPP/05/20 \\
DCPT/05/40 \\
MAN/HEP/2005/2 \\
12 December 2005 \\
\end{flushright}

{\par \noindent \textbf {\large Detecting the Standard Model Higgs Boson in the WW decay channel using forward proton tagging at the LHC}\large \par}
\bigskip{}
{\par \centering \it B. E. Cox$^{a}$, A. De ~Roeck$^{b}$, V.A. ~Khoze$^{c}$, T. Pierzchala$^{d}$, M.G. Ryskin$^{c,e}$, 
I. Nasteva$^{a}$, W. J. Stirling$^{c}$ and M. Tasevsky$^{f}$} \\
\bigskip{}
{\par \noindent $^a$ Department of Physics and Astronomy, University of Manchester,
Manchester M139PL, UK \\
$^b$ CERN,1211,Geneva 23, Switzerland \\
$^c$ Department of Physics and Institute for
Particle Physics Phenomenology, University of Durham,\\ DH1 3LE, UK \\
$^d$ UCL-FYNU, D\'{e}partement de physique, Universit\'{e} catholique
de Louvain, B-1348 Louvain-la-Neuve, Belgie \\
$^e$ Petersburg Nuclear Physics Institute, Gatchina, St. Petersburg, 188300, Russia \\
$^f$ University of Antwerp, Physics Department, 2610 Antwerp, 
Belgium \\
}
\begin{abstract}
\noindent
We present a detailed study of the central exclusive production of the Standard Model Higgs Boson in the $WW$ decay channel at the LHC. We include estimates of the experimental acceptance, including that of the proposed proton tagging detectors at 220m and 420m around either ATLAS and / or CMS, and the level 1 trigger acceptances. We give first estimates of the photon-photon and glue-glue background processes in the semi-leptonic and fully-leptonic decay channels. We find that there will be a detectable signal for Higgs masses between 140 GeV and 200 GeV, and that the backgrounds should be controllable.

\end{abstract}

\end{titlepage}

% ===========================================================================
\section{Introduction}

The use of forward proton tagging as a means to discover new physics at the LHC has received a great deal of 
attention recently (see for example \cite{Albrow:2000na,DeRoeck:2002hk,Piotrzkowski:2000rx,Assamagan:2004mu,Boonekamp:2004nu,Cox:2004rv,Kisselev:2005tp,Khoze:2005ie,Forshaw:2005qp} and references therein). 
The process of interest is the so-called `central exclusive' process, $pp\rightarrow p \oplus \phi \oplus p$, 
where $\oplus$ denotes the absence of hadronic activity (`gap') between the outgoing protons and the 
{\rm }decay products of the central system $\phi$. There are two primary reasons, from which all other advantages follow, 
that central exclusive production is attractive. Firstly, if the outgoing protons remain intact and scatter
 through small angles, then, to a very good approximation, the central system $\phi$ is produced in the $J_z=0$,
 C and P even state. An absolute determination of the quantum numbers of any resonance is possible by measurements of the correlations between outgoing proton momenta.
Secondly, the mass of the central system can be determined very accurately from a measurement 
of the transverse and longitudinal momentum components of the outgoing protons alone. For the case of exclusive particle production, 
this means an accurate determination of the mass irrespective of the decay mode of the centrally produced 
particle. 

There are several locations around the LHC interaction points at which it is possible to install forward proton tagging
detectors. The 220m region will almost certainly 
be instrumented at both ATLAS and CMS at LHC startup \cite{totemtdr,atlasltdr}, and there 
are plans to install tagging detectors in the 420m region at some point in the future \cite{FP420}.   
Recent studies suggest that the missing mass resolution will be $\sigma \sim 1\%$ for a $140$ GeV central system, if both
protons are detected at 420m from the interaction point \cite{ristoetal}. For configurations in which one proton 
is detected at 220m, and one at 420m, the resolution deteriorates to approximately $6\%$. There is no acceptance for 
central systems with masses less than $\sim 200$ GeV with 220m detectors alone. 

Previous analyses have focused primarily on light ($115$ GeV $< M < 160$ GeV) Standard Model Higgs production, with the Higgs 
decaying to 2 b-jets \cite{DeRoeck:2002hk}. The potentially copious b-jet background is controlled by a combination of the $J_z=0$
selection rule, which 
strongly suppresses central $b \bar b$ production at leading order \cite{Khoze:2000jm}, and the mass resolution from the tagging 
detectors. The missing mass resolution is critical to controlling the background because the remaining b-jet background is a 
continuum beneath the Higgs mass peak, and therefore poor resolution simply allows 
more background events into the mass window around the peak. Assuming a Gaussian mass resolution of width $\sigma = 1$ GeV, 
it is estimated that 
11 signal events, with a signal to background ratio of order 1, can be achieved with a luminosity of 30~fb$^{-1}$ in the $b \bar b$ decay
channel \cite{DeRoeck:2002hk}. It is worth noting that in the large tan $\beta$ region of MSSM parameter space, 
the situation becomes much more 
favourable, leading to predicted signal to background ratios in excess of 20 for the lightest Higgs mass of 
$\sim 130$ GeV \cite{Kaidalov:2003ys}. The central exclusive channel may be the discovery channel in this case.

Whilst the $b \bar b$ channel is certainly attractive, since it allows direct access to the dominant decay mode of the light Higgs, 
 there are two potential problems which render it challenging from an experimental perspective.
Firstly, since the mass resolution of the proton taggers is used to suppress the background, any degradation in the  
expected resolution will adversely 
affect the signal to background ratio. Secondly, level 1 triggering of $H \rightarrow b \bar b$ events is difficult. The 
420m detectors are at or beyond the distance at which signals arrive at the central detectors in time for a level 1 trigger decision. Triggering on the 
central system may therefore be necessary, but the low-mass di-jet signature of the $H \rightarrow b \bar b$ channel will certainly be a challenge for 
both ATLAS and CMS. 

In this paper, we turn our attention to the $WW^*$ decay mode of the light Standard Model Higgs Boson, and above the 2 $W$ threshold, 
the $WW$ decay mode. As we shall see, this channel 
does not suffer from either of the above problems: suppression of the dominant backgrounds does not rely primarily on the mass resolution of the 
detectors, and certainly in the leptonic and semi-leptonic decay channels, level 1 triggering is not a problem. The advantages of forward proton tagging are, however, 
still explicit. Even for the double leptonic decay channel (i.e. two final state neutrinos), 
the mass resolution should be better than $6 \%$ (and $\sim 1 \%$ for double 420m-tagged events), and is expected to improve further 
with increasing Higgs mass, and of course 
observation of the Higgs in the exclusive double tagged channel immediately establishes its quantum numbers.  

In section \ref{signal} we use the ExHuME Monte Carlo \cite{Monk:2005ji}
to simulate the signal process $pp \rightarrow p \oplus H \oplus p \rightarrow p \oplus WW \oplus p$, and discuss possible  
trigger strategies for the LHC experiments. ExHuME is a direct implementation of the calculations of 
\cite{Khoze:2000cy,Khoze:2001xm}. In section \ref{background} we survey the backgrounds, and in section 
\ref{sb} we present our conclusions.     
\begin{figure}[htb]        
\begin{center}
  \setlength{\unitlength}{1 mm}      
  \Large 
\begin{picture}(140,120)(0,0)
    \put(0,0){\epsfig{file=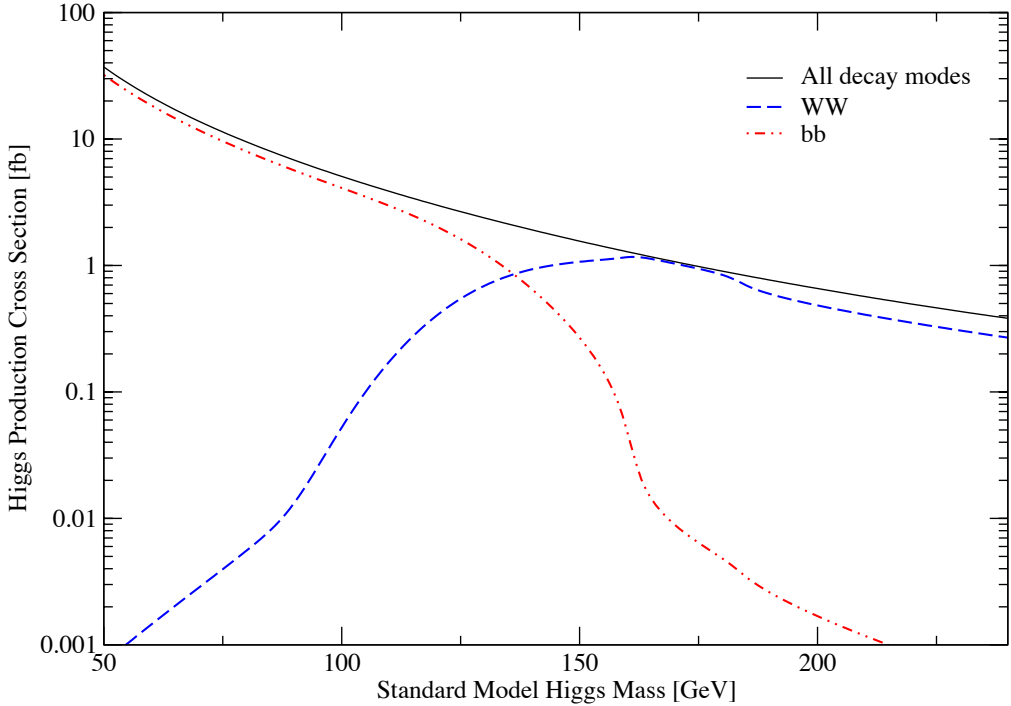,width=1.\textwidth}}
\end{picture}
\end{center}
\caption 
        {The cross section times branching ratio for the central exclusive production of the Standard Model Higgs boson as a function of Higgs mass in the $WW$ and $b \bar b$ decay channels.\label{wwxs}}

\end{figure}

\section{The Signal}
\label{signal}

The central exclusive production cross section for the Standard Model Higgs Boson was calculated in 
\cite{Khoze:2000cy,Khoze:2001xm}.  
In figure \ref{wwxs} we show the cross section for the process $pp \rightarrow p H p \rightarrow p WW p$ as a function of the Higgs mass $M_H$. The increasing 
branching ratio to $WW$ as $M_H$ increases
compensates for the falling central exclusive production cross section. For comparison we also show the cross 
section times branching ratio for $pp \rightarrow p H p \rightarrow p b \bar b p$. 
We also expect the $WW$ channel to be effective in the study of the low tan $\beta$ MSSM \cite{KRS}, although we leave 
this for a future publication. 

Events with two $W$ bosons in the final state fall into 3 broad categories from an experimental perspective, depending on 
the decay modes of the $W$. Events in which at least one of the $W$ bosons decays in either the $e$ or $\mu$ channel 
are the simplest, and will usually pass the level 1 trigger thresholds of ATLAS and CMS due to the high $p_T$ final state lepton, as we shall see below.  
If neither of the $W$ bosons decay in the $e$ or $\mu$ channel, the event can still pass the level 1 trigger thresholds if a $W$ decays in the $\tau$ channel, with 
the $\tau$ subsequently decaying leptonically (although the leptons from the $\tau$ decays have a softer $p_T$ spectrum).
The 4-jet decay mode occurs approximately half the time, but it is unlikely that this signature will pass the level 1 triggers 
of either ATLAS or CMS without information from the proton taggers. It is possible that the 220m proton detectors can be included in the level 1 trigger and therefore events with one proton detected at 220m could be taken, even in the 4-jet case. This will also increase the trigger efficiency in the leptonic and semi-leptonic channels. We leave this possibility for future study.

  \begin{figure}[htb]        
\begin{center}
  \setlength{\unitlength}{1 mm}      
  \Large 
\begin{picture}(160,70)(0,0)
    \put(0,0){\epsfig{file=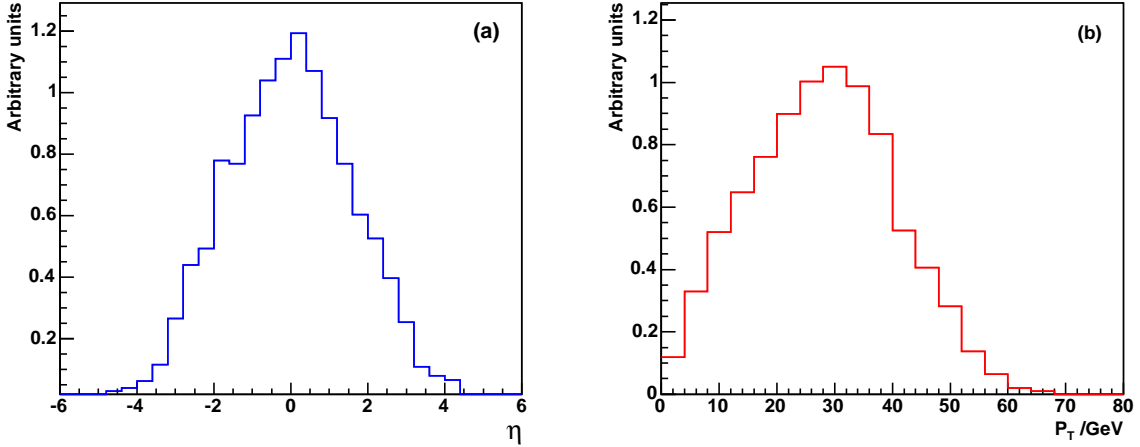,width=1.\textwidth}}
\end{picture}
\end{center}
\caption 
        {The pseudorapidity, $dN / d \eta$, (a) and transverse momentum, $dN / d p_T$, (b) distributions of the highest $p_T$ lepton for $M_H = 140$ GeV in the process $pp \rightarrow p H p \rightarrow p WW p$ \label{wwleptons}}

\end{figure}
\begin{figure}[htb]        
\begin{center}
  \setlength{\unitlength}{1 mm}      
  \Large 
\begin{picture}(160,70)(0,0)
    \put(0,0){\epsfig{file=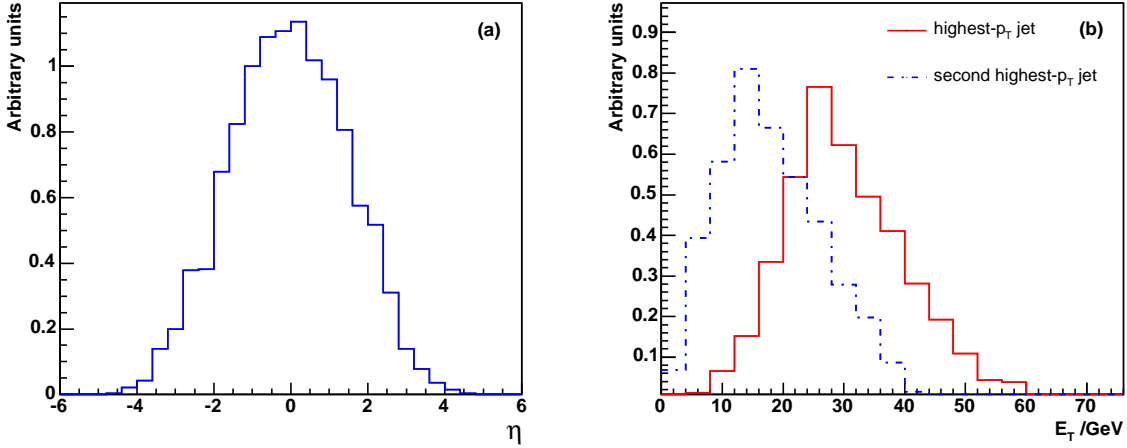,width=1.\textwidth}}
\end{picture}
\end{center}
\caption 
        {The pseudorapidity, $dN / d \eta$, and transverse momentum, $dN / d p_T$, distributions of the 2 highest $p_T$ jets for $M_H = 140$ GeV. Details of the jet finding algorithm can be found in the text \label{wwjets}}
\end{figure}

 
In figure \ref{wwleptons} we show the pseudorapidity and transverse momentum distributions of the decay lepton in events 
where at least one $W$(or a $\tau$ lepton in the case of the $W \rightarrow \tau \nu_{\tau}$ decay mode)
 decays in the $e$ or $\mu$ channel, for $M_H = 140$ GeV. For the doubly leptonic decay modes, the highest transverse momentum 
lepton is chosen. Events were generated using ExHuME 1.3 \cite{Monk:2005ji} interfaced 
to PYTHIA 6.205 \cite{Sjostrand:2000wi} for the decay of the Higgs boson. 
The CMS level 1 trigger has a single electron threshold of $29$ GeV with a pseudorapidity coverage of $|\eta| < 2.5$,
 and $14$ GeV for a single muon with $|\eta| < 2.5$ \cite{cmstdr}. At ATLAS, the level 1 trigger thresholds  
are $25$ GeV, 
$|\eta| < 2.5$ for single electrons and $20$ GeV, $|\eta| < 2.5$ for single muons \cite{atlastdr}. From figure \ref{wwleptons}, it is clear 
that a reasonable fraction of signal 
events will be taken by these standard triggers. In figure \ref{wwjets} we show the jet $E_T$ and pseudorapidity distributions
 (2 entries per event) for the two highest $p_T$ jets in the semi-leptonic decay channel. The jets are found using the exclusive
$k_T$ algorithm \cite{Butterworth:2002xg}, in the $E$ scheme \footnote{The `E' scheme means that the jet 4-vectors are formed by 4-vector 
addition of the particles that make up the jet}. In exclusive mode, the final state is forced into a 2-jet topology. The merging scale 
(often termed $y_{cut}$) which defines the two jets has been used as a powerful background suppression tool 
(see for example \cite{Butterworth:2002tt}). We do not simulate the background processes to the hadronic final state level  
in this paper, and 
we therefore leave the details of optimising the jet finding for a later publication, whilst noting that such optimisation will
undoubtedly be important. Requiring $2$ central jets (and perhaps reduced hadronic activity outside the jets) at level 1 in conjunction with a high $p_T$ electron or muon should allow the single lepton trigger 
thresholds to be further reduced. It is also likely that the trigger efficiency in the $\tau$ channel can be improved. 
For the purposes of this paper, events with $\tau$ decays are kept only if they pass the standard trigger definitions outlined above. In summary, our trigger efficiency estimates here are conservative.    

  Besides the level 1 trigger, the other important ingredient in the efficiency comes from the acceptance of the proton taggers themselves. The acceptances were obtained using a fast simulation program for the CMS detector, which includes a parameterisation of the response of the Roman Pots based on a detailed simulation \cite{famos} where the scattered protons were tracked with the MAD \cite{madx} package. MAD was used with LHC optics version 6.2. The acceptance rises from 60\% at $M_H = 120$ GeV to 80\% at $M_H = 200$ GeV.      
 
\begin{table}
\centering \begin{tabular}{|c|c|c|c|c|}
\hline
 Selection cuts& Higgs Mass& Efficiency& $\,$ Signal $\,$ & Events \\
  & (GeV)& & \( \sigma \) (fb)& / 30 fb$^{-1}$ \\
\hline
  & 120 &  100\%&  0.403 & 12.1 \\
\textbf{Generated}& 140& 100\% & 0.933 & 28.0 \\
  $H \rightarrow WW$ &160 &  100\% & 1.164 & 34.9 \\
  & 180&  100\% & 0.843 & 25.3 \\
  & 200 & 100\% & 0.483 & 14.5 \\
\hline
  & 120 &   61 \%  & 0.246 & 7.4 \\
Acceptance of proton taggers & 140 &  67 \%  & 0.625 & 18.8 \\
(420m + 220m) & 160 &  71 \% & 0.826 & 24.8 \\
 & 180 &   74 \% & 0.624 & 18.7 \\
 & 200 &   77 \% & 0.372 & 11.2 \\
\hline\hline
 Single lepton trigger:& 120 & 8.7 \%& 0.035 & 1.1 \\
 an electron with \( p_{T} \)\( > 25 \) GeV & 140&  12.8 \% & 0.119 & 3.6 
\\
 or a muon with \( p_{T} \)\( > 20 \) GeV& 160 &  16.6 \% & 0.194 & 5.8 \\
 within \( |\eta| < 2.5 \) & 180&  18.3 \% & 0.154 & 4.6 \\
 & 200 & 19.8 \% & 0.096 &  2.9 \\
\hline
 & 120& 7.0 \% & 0.028 & 0.8 \\
 2 or more jets & 140&  10.2 \% & 0.096 & 2.9  \\
 within  \( |\eta| < 2.5 \)& 160 & 13.6 \% & 0.158 & 4.7  \\
 & 180& 15.1 \% & 0.127 & 3.8 \\
 & 200& 16.6 \% & 0.080 & 2.4 \\
\hline
& 120& 0.54 \% & 0.002 & 0.1 \\
 Mass window around & 140& 2.0 \% & 0.019 & 0.6 \\
 hadronically decaying $W$ & 160&  7.2 \% & 0.084 & 2.5 \\
 70 GeV \( < M_W < \) 90 GeV& 180&  9.5 \% & 0.080  & 2.4 \\
 & 200&  10.8 \% & 0.052 & 1.6 \\
\hline
 & 160&  6.6 \% & 0.077 & 2.3 \\
 \( p_T \)(protons) \( > 100 \) MeV & 180&  8.6 \% & 0.073 & 2.2 \\
 & 200&  9.8 \% & 0.047 & 1.4 \\
\hline
 & 160&  5.2 \% & 0.061 & 1.8 \\
 \( p_T \)(protons) \( > 200 \) MeV & 180&  6.7 \% & 0.057 & 1.7 \\
 & 200&  7.7 \% & 0.037 & 1.1 \\
\hline
\end{tabular}\small \par

\caption{\label{signal1}The effect of cuts on signal samples for selecting 
semileptonic WW decays 
($WW \to l\nu jj$, $l=e,\mu, \tau$, $\tau \to e,\mu$) 
for different Higgs masses using the standard ATLAS leptonic trigger 
thresholds.}
\end{table}

In table 1 we show the efficiency for detection of the semi-leptonic decay channel, and in table 2, the fully leptonic decay channel. We use the ATLAS trigger thresholds as an example. The CMS thresholds give similar results. It is worth stressing again that we expect it to be possible to significantly improve the trigger efficiencies quoted in line 3 of tables 1 and 2. For example, reducing the single $e$ and $\mu$ trigger thresholds to $15$ GeV would 
increase the trigger efficiency by $\sim 50$ \%.

\begin{table}
\centering \begin{tabular}{|c|c|c|c|c|}
\hline
 Selection cuts& Higgs Mass& Efficiency& $\,$ Signal $\,$ & Events \\
 &(GeV) &  & \( \sigma \) (fb) & / 30 fb$^{-1}$ \\
\hline
  & 120 &  100\%&  0.403 & 12.1 \\
\textbf{Generated}& 140& 100\% & 0.933 & 28.0 \\
  $H \rightarrow WW$ &160 &  100\% & 1.164 & 34.9 \\
  & 180&  100\% & 0.843 & 25.3 \\
  & 200 & 100\% & 0.483 & 14.5 \\
\hline
  & 120 &   61 \%  & 0.246 & 7.4 \\
Acceptance of proton taggers & 140 &  67 \%  & 0.625 & 18.8 \\
(420m + 220m) & 160 &  71 \% & 0.826 & 24.8 \\
 & 180 &   74 \% & 0.624 & 18.7 \\
 & 200 &   77 \% & 0.372 & 11.2 \\
\hline\hline
Single and di-lepton triggers: & 120 &   2.3 \%  & 0.009 & 0.3 \\
$2e$ ($p_{T}^{e}>15$ GeV) or $2\mu$ ($p_{T}^{\mu}>10$ GeV) & 140 &  3.1 \%  
& 0.029 & 0.9 \\
or $2e$ ($p_{T,max}^{e}>25$ GeV) or $2\mu$ ($p_{T,max}^{\mu}>20$ GeV) & 
160 &  3.3 \% & 0.038 & 1.2 \\
or $e\mu$ ($p_{T}^{e}>15$ GeV and $p_{T}^{\mu}>10$ GeV) & 180 &  3.5 \% & 
0.030 & 0.9 \\
or $e\mu$ ($p_{T}^{e}>25$ GeV or $p_{T}^{\mu}>20$ GeV) & 200 &  3.6 \% & 
0.017 & 0.5 \\
within \( |\eta| < 2.5 \) &  & &  &  \\
\hline
 & 160 &   3.1 \% & 0.036 & 1.1 \\
\( p_T \)(protons) \( > 100 \) MeV & 180 &   3.2 \% & 0.027 & 0.8 \\
 & 200 &   3.3 \% & 0.016 & 0.5 \\
\hline
 & 160 &   2.4 \% & 0.028 & 0.8 \\
\( p_T \)(protons) \( > 200 \) MeV & 180 &   2.5 \% & 0.021 & 0.6 \\
 & 200 &   2.5 \% & 0.011 & 0.3 \\
\hline
\end{tabular}\small \par    
\caption{\label{signal2}The effect of cuts on signal samples for selecting 
fully leptonic WW decays 
($WW \to l\nu l\nu$, $l=e,\mu,\tau$, $\tau \to e,\mu$) 
for different Higgs masses using the standard ATLAS 
single and double leptonic trigger thresholds.}
\end{table}

\section{The Backgrounds}
\label{background}
\begin{figure}[htb]        
\begin{center}
  \setlength{\unitlength}{1 mm}      
  \Large 
\begin{picture}(160,120)(0,0)
    \put(0,0){\epsfig{file=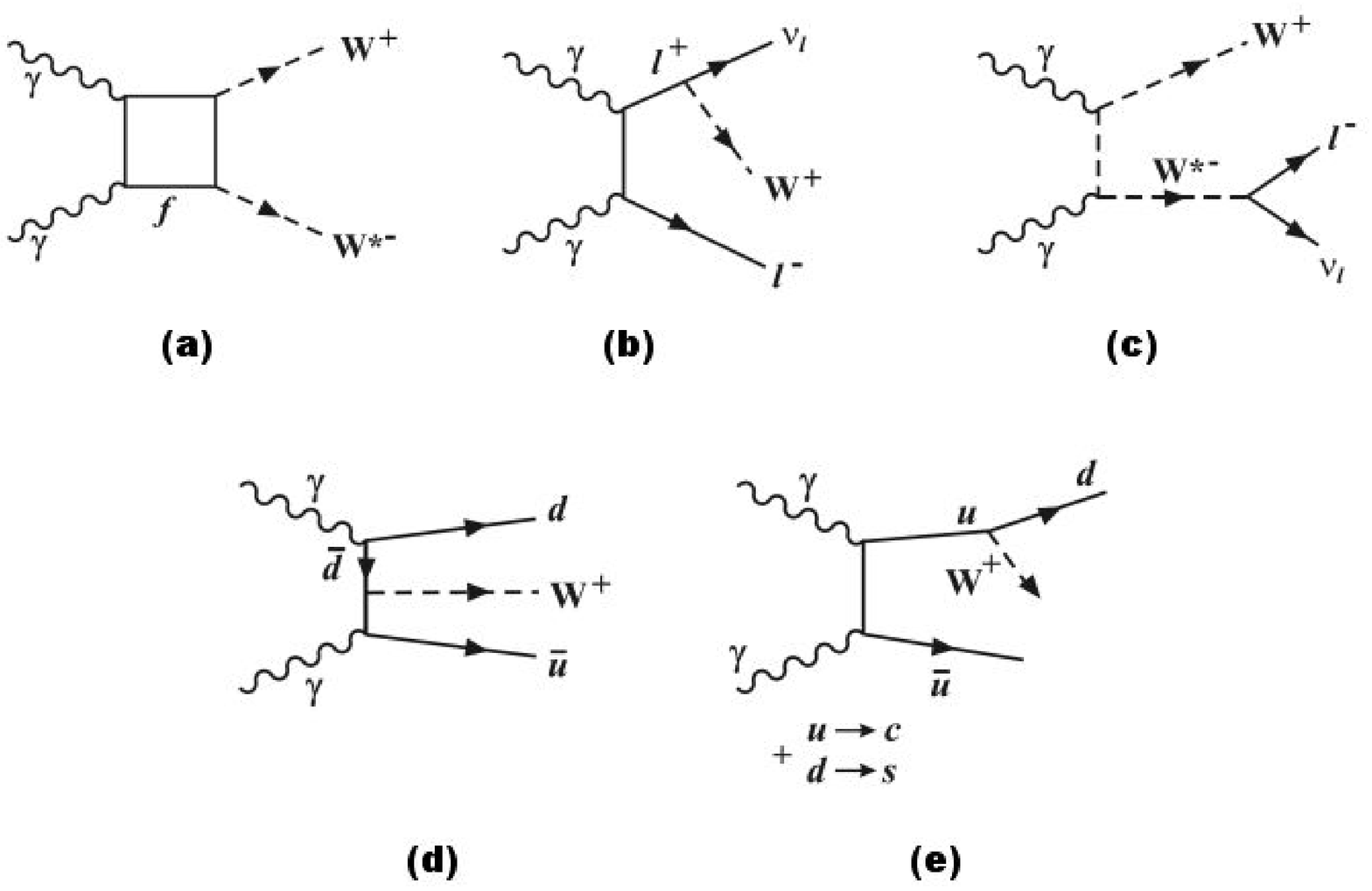,width=1.\textwidth}}
\end{picture}
\end{center}
\caption 
        {Photon-induced background processes.\label{bg1}}

\end{figure}
\begin{figure}[htb]        
\begin{center}
  \setlength{\unitlength}{1 mm}      
  \Large 
\begin{picture}(160,55)(0,0)
    \put(0,0){\epsfig{file=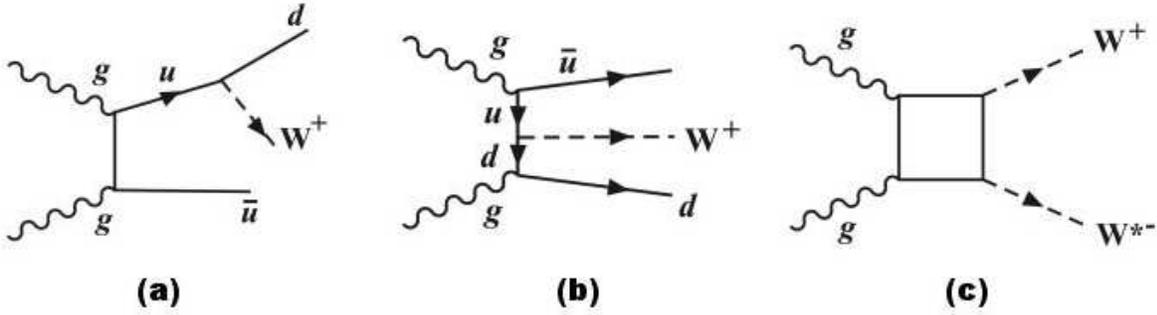,width=1.\textwidth}}
\end{picture}
\end{center}
\caption 
        {Gluon-induced background hard sub-processes. \label{bg2}}

\end{figure}

 
One of the attractive features of the $WW$ channel is the lack of a relatively large 
irreducible continuum background process, such as 
central exclusive $b \bar b$ production in the case of $H \rightarrow b \bar b$, which 
relies on the experimental missing mass resolution being good enough to provide adequate suppression. 
The primary  exclusive  backgrounds 
are illustrated in figures \ref{bg1} and \ref{bg2}. Tree-level photon-photon processes (figures 4 (b-e)) are calculated using CalcHEP \cite{Pukhov:1999gg,Pukhov:2004ca}.
The background 
coming from processes of the type shown in figures \ref{bg1} (b), (d) and (e) is potentially large due to collinear 
logarithms corresponding to lepton (or quark) production at small angles. By requiring that the leptons (or jets) are central the contribution from such diagrams decreases. After imposing the pseudorapidity cut $|\eta| < 2.5$, the cross sections corresponding to all tree-level photon-induced subprocesses are
 $\sigma^{QED}(pp \rightarrow  pWWp)$ =  0.015 (0.033, 0.37, 2) fb for $M_H$ = 120 (140, 160, 180) GeV where we
integrated over the mass interval $\Delta M\simeq 3\sigma\sim
  0.05 M_H$, assuming a Gaussian mass resolution 
of the proton taggers of width of $2$ GeV \footnote{We note that the resolution for events in which both protons are in the 420m taggers may be better than 2~GeV, and events where one proton is detected at 220m and one at 420m may be worse. We take this figure as a plausible example}. Note that these cross sections include a gap survival factor,
which according to the calculations of \cite{KMRphot}  is  $S^2 \simeq 0.9$ for the integrated cross section.
Therefore, even without additional experimental cuts on the final state, this background contribution is comfortably below the signal 
cross section for $M_H < 150$ GeV. After applying the single leptonic trigger cuts as detailed in table 1, the QED background cross sections become
$\sigma^{QED}=$ 0.01 (0.02,0.27,1.53) fb for $M_H =$ 120 (140,160,180) GeV
The overall photon-photon background contribution rises
with $M_H$. 
Imposing a cut on the
transverse momenta of the outgoing protons $p_T>
100 (200)$ MeV suppresses the photon fusion by process approximately a factor of $15 (75)$,
whilst reducing the signal by 10\% (40 \%)\footnote{Note that minimum $p_T$ cuts on the final-state protons induce 
a further reduction of the survival factor
in QED-induced processes, for example for  $p_T> 100 (200)$~MeV  $S^2 \simeq 0.6 (0.5)$}.
 Such a cut will most likely be necessary 
for Higgs masses above the WW threshold, and we include it for large Higgs masses in tables \ref{signal1} and 
\ref{signal2} \footnote {We note that a 100 MeV cut on the  $p_T$ of the outgoing protons
is close to the intrinsic $p_T$ spread of the LHC beams, and it may therefore be necessary
to raise this cut in practice \cite{Piotrzkowski:2000rx}}.. Further optimisation of the cuts on the final-state particles should enable
this QED background to be reduced further without dramatically affecting the signal. In principle, the angular
distributions and correlations between the reconstructed $W$ bosons will be different for the (scalar) Higgs decay and 
the photon fusion backgrounds. With the expected low number of signal events and the centrality requirement on the $W$
decay products however, such techniques are unlikely to be useful.    

\begin{figure}[htb]        
\begin{center}
  \setlength{\unitlength}{1 mm}      
  \Large 
\begin{picture}(160,120)(0,0)
    \put(0,0){\epsfig{file=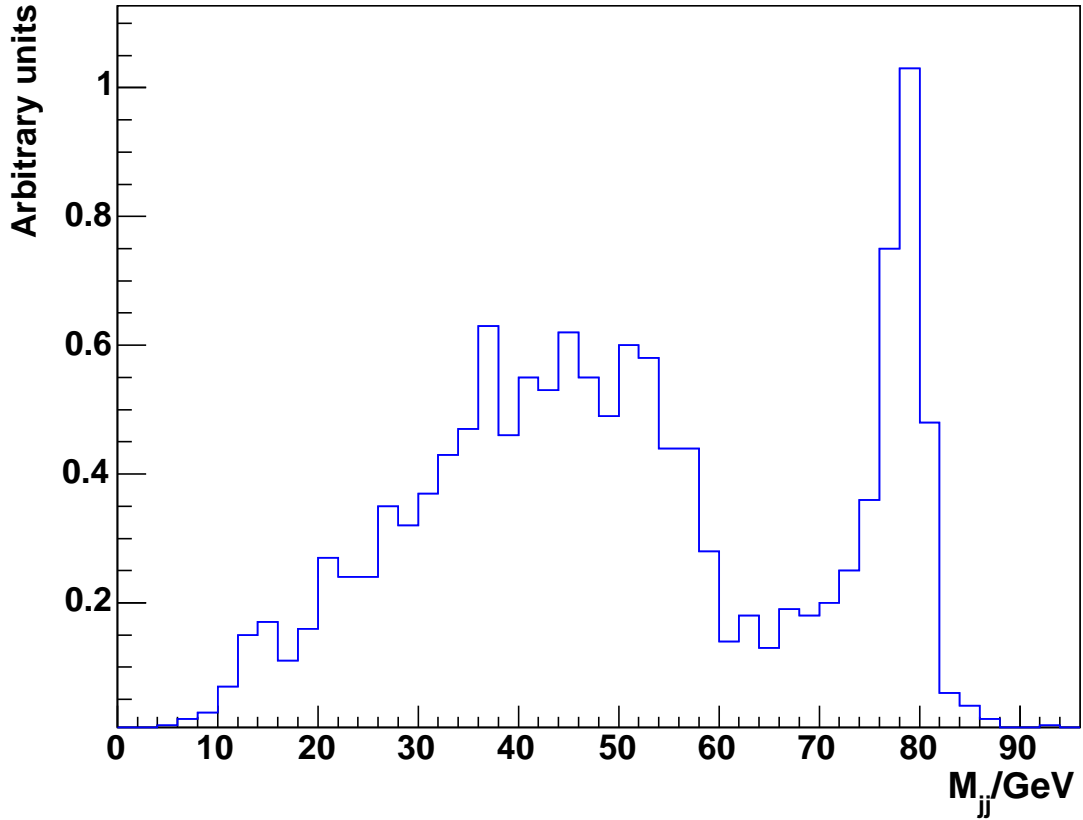,width=1.\textwidth}}
\end{picture}
\end{center}
\caption 
        {The di-jet invariant mass distribution $dN / dM_{jj}$ in the semi-leptonic decay 
channel $H \to WW^* \to l \nu j j $ for $M_H = 140$ GeV.\label{mjj}}

\end{figure}

The other important background comes from the QCD $W$-strahlung
sub-processes of the type shown in figures 5 (a) and (b) \footnote{In figure 5 the intact protons and screening gluon are omitted for clarity}, which have recently been studied in \cite{KRS} 
\footnote{The contribution of the box diagram figure 5 (c) is much smaller, as can be deduced from the 
results of Ref.~\cite{Duhrssen:2005bz}}.  
Here
we have to take into account the non-trivial polarization structure of the
$J_z=0$ amplitude. This was done in \cite{KRS} using the spinor technique of
\cite{Kleiss:1985yh}.
Again to suppress the collinear (quark line) logarithms we impose
the pseudorapidity cut $|\eta_{jet}|<2.5$ on the final state quarks.
With these kinematic cuts, the cross section for the processes in figures 5 (a) and (b), summed over the two
families of fermions and including both $W^+W^{*-}$ and $W^-W^{*+}$
configurations, is $7.2$ $(9.7)$ pb for $M_H=120 (140)$ GeV. It
rises to $10.6$ pb at $M_H=160$ GeV and then decreases slowly for higher masses, falling to $9.5$ pb at $M_H = 200$ GeV. 
This cross section
should be multiplied by the phase space factor $2\Delta M/M_H\sim 0.1$, again 
assuming  $\Delta M\simeq 3\sigma\sim 0.05 M$, 
and by the corresponding gluon luminosity \cite{Khoze:2001xm}. This leads
to the QCD background cross section at $\sqrt{s}=14$ TeV of 1.7 fb for
$M_H=140$ GeV. Above the 2-W threshold, we expect to be able to suppress this background extremely effectively by requiring that 
the quark jets fall within an appropriate $W$ mass window, since the background will be a continuum beneath the $W$ mass peak. 
The potential problem is therefore only for Higgs masses below 160 GeV, for the case in which the $W^*$ decays hadronically 
and therefore the di-jet mass from the signal events falls outside the $W$ mass window. In Figure \ref{mjj} we show the di-jet mass 
distribution for the semi-leptonic W-decay channel, $M_H=140$ GeV, after the level 1 leptonic trigger and di-jet pseudorapidity cuts, for the signal sample only.
Imposing the $W$ mass window effectively removes all hadronically decaying $W^*$ events from 
the sample, with the benefit of greatly enhancing the signal to background ratio. As can 
be seen from table 1, this reduces the signal by a factor of $\sim 5$ for $M_H = 140$ GeV.  
A fraction of these lost events can be recovered if the leptonic trigger thresholds are reduced at level 1, since forcing the $W^*$ to 
decay leptonically reduces the average $p_T$ of the decay leptons. There may be alternative ways of reducing this background, for example by imposing cuts on 
the final state to account for the azimuthal correlations between the quark jets. A full Monte Carlo simulation of these QCD processes will be 
required to assess the effectiveness of such approaches.

Finally, we consider the fully leptonic decay modes. The only background subprocesses in this case are caused by the photon fusion diagrams of figures 4 (a), (b) and (c) and by the QCD box diagram of figure 5 (c). As we discuss above, the photon-induced contributions can be reduced in the high-mass region, where they become potentially large, by the proton transverse momentum cut $p_T > 100$ MeV. Using the results of \cite{Duhrssen:2005bz} we find the QCD contribution from figure 5 (c) is very small, less than 1\% of the signal.

To summarise, the most problematic background contribution arises in the semi-leptonic case from the QCD diagrams shown in figures 5 (a) and (b) when the off-shell $W$ boson decays hadronically. At Higgs masses above $150$ GeV, photon-induced backgrounds from diagrams of the type shown in figure 4 can be a problem, but can be suppressed if necessary by increasing the transverse momentum cut on the tagged protons to $p_T > 200$ MeV. As shown in tables 1 and 2, this higher $p_T$ cut has little effect on the signal, but will further reduce the photon-induced backgrounds. 
For the fully leptonic decay modes, and for semi-leptonic 
decays in which the on-mass-shell $W$ boson decays hadronically, the signal to background ratio should be much greater than unity. We have not considered the $ZZ$ decay channel here, because we expect the rate to be too low to be of interest.

\section{Conclusions}
\label{sb}

We have shown that, given the standard level 1 trigger thresholds at both ATLAS and CMS, and installation of the proposed 220m and 420m proton tagging detectors, we expect the Standard Model Higgs boson to be visible in the $WW / WW^*$ double tagged exclusive channel for $140$ GeV $< M_H < 200$ GeV with $30$~fb$^{-1}$ of LHC luminosity. For a 120 GeV Higgs in the $WW^*$ channel, the event yield is marginal. For masses above 140 GeV, we expect approximately 5 or 6 events, largely independent of Higgs mass, of which 1 is expected to be in the `gold plated' doubly leptonic channel, with no appreciable background. These numbers would double if the trigger thresholds on single leptons could be reduced to $15$ GeV, which may be possible by using other event characteristics such as the $2$ central jets from the hadronically decaying $W$. When a di-jet 
$W$ mass window is applied, 2 or 3 events remain, with again no appreciable backgrounds. In the semi-leptonic channel, there is a potentially dangerous background from central exclusive $W$ + jets processes below the 2-$W$ threshold, although we expect that this background may be manageable with carefully chosen experimental cuts. The largest loss of events is caused by the level 1 trigger efficiency, and we have made no attempt to optimise this here, although we expect that significant improvements will be possible.

\section*{Acknowledgments}
We thank J.~Forshaw, H. Kowalski, A.D.~Martin, J.~Monk, A.~Nikitenko, R.~Orava, A.~Pilkington and K.~Piotrzkowski for stimulating discussion. This 
work was supported in part by grant RFBR 04-02-16073, by the UK Particle Physics and Astronomy Research Council, and by the Belgian Interuniversity Attraction Poles Programme.

% ========================================================================


\newpage
\begin{thebibliography}{xx}

%\cite{Albrow:2000na}
\bibitem{Albrow:2000na}
  M.~G.~Albrow and A.~Rostovtsev,
  %``Searching for the Higgs at hadron colliders using the missing mass
  %method,''
  arXiv:hep-ph/0009336.
  %%CITATION = HEP-PH 0009336;%%

%\cite{DeRoeck:2002hk}
\bibitem{DeRoeck:2002hk}
A.~De Roeck, V.~A.~Khoze, A.~D.~Martin, R.~Orava and M.~G.~Ryskin,
%``Ways to detect a light Higgs boson at the LHC,''
Eur.\ Phys.\ J.\ C {\bf 25} (2002) 391
[arXiv:hep-ph/0207042].
%%CITATION = HEP-PH 0207042;%%

%\cite{Piotrzkowski:2000rx}
\bibitem{Piotrzkowski:2000rx}
K.~Piotrzkowski,
%``Tagging two-photon production at the LHC,''
Phys.\ Rev.\ D {\bf 63} (2001) 071502
[arXiv:hep-ex/0009065].
%%CITATION = HEP-EX 0009065;%%

%\cite{Assamagan:2004mu}
\bibitem{Assamagan:2004mu}
  K.~A.~Assamagan {\it et al.}  [Higgs Working Group Collaboration],
  %``The Higgs working group: Summary report,''
  arXiv:hep-ph/0406152.
  %%CITATION = HEP-PH 0406152;%%

%\cite{Boonekamp:2004nu}
\bibitem{Boonekamp:2004nu}
M.~Boonekamp, R.~Peschanski and C.~Royon,
%``Sensitivity to the standard model Higgs boson in exclusive double
%diffraction,''
Phys.\ Lett.\ B {\bf 598} (2004) 243
[arXiv:hep-ph/0406061].
%%CITATION = HEP-PH 0406061;%%

%\cite{Cox:2004rv}
\bibitem{Cox:2004rv}
B.~E.~Cox,
%``A review of forward proton tagging at 420-m at the LHC, and relevant results
%from the Tevatron and HERA,''
AIP Conf.\ Proc.\  {\bf 753} (2005) 103
[arXiv:hep-ph/0409144].
%%CITATION = HEP-PH 0409144;%%

%\cite{Kisselev:2005tp}
\bibitem{Kisselev:2005tp}
A.~V.~Kisselev, V.~A.~Petrov and R.~A.~Ryutin,
%``5-dimensional quantum gravity effects in exclusive double diffractive
%events,''
arXiv:hep-ph/0506034.
%%CITATION = HEP-PH 0506034;%%

%\cite{Khoze:2005ie}
\bibitem{Khoze:2005ie}
V.~A.~Khoze, A.~B.~Kaidalov, A.~D.~Martin, M.~G.~Ryskin and W.~J.~Stirling,
%``Diffractive processes as a tool for searching for new physics,''
arXiv:hep-ph/0507040.
%%CITATION = HEP-PH 0507040;%%

%\cite{Forshaw:2005qp}
\bibitem{Forshaw:2005qp}
J.~R.~Forshaw,
%``Diffractive Higgs production: Theory,''
arXiv:hep-ph/0508274.
%%CITATION = HEP-PH 0508274;%%

\bibitem{totemtdr}
TOTEM Collaboration, Technical Design Report, CERN/LHCC/2004-002

\bibitem{atlasltdr}
ATLAS Collaboration, Luminosity Letter of Intent, CERN/LHCC/2004-010

\bibitem{ristoetal}
T. M\"aki et al.,to be published in the proceedings of the 'HERA and the LHC' Workshop

%\cite{Khoze:2000jm}
\bibitem{Khoze:2000jm}
V.~A.~Khoze, A.~D.~Martin and M.~G.~Ryskin,
%``Double-diffractive processes in high-resolution missing-mass  experiments at
%the Tevatron,''
Eur.\ Phys.\ J.\ C {\bf 19} (2001) 477
[Erratum-ibid.\ C {\bf 20} (2001) 599]
[arXiv:hep-ph/0011393].

%\cite{Kaidalov:2003ys}
\bibitem{Kaidalov:2003ys}
A.~B.~Kaidalov, V.~A.~Khoze, A.~D.~Martin and M.~G.~Ryskin,
%``Extending the study of the Higgs sector at the LHC by proton tagging,''
Eur.\ Phys.\ J.\ C {\bf 33} (2004) 261
[arXiv:hep-ph/0311023].
%%CITATION = HEP-PH 0311023;%%
%\cite{Monk:2005ji}
\bibitem{Monk:2005ji}
J.~Monk and A.~Pilkington,
%``ExHuME: A Monte Carlo Event Generator for Exclusive Diffraction,''
arXiv:hep-ph/0502077.
%%CITATION = HEP-PH 0502077;%%

%\cite{Khoze:2000cy}
\bibitem{Khoze:2000cy}
  V.~A.~Khoze, A.~D.~Martin and M.~G.~Ryskin,
  %``Can the Higgs be seen in rapidity gap events at the Tevatron or the
  %LHC?,''
  Eur.\ Phys.\ J.\ C {\bf 14} (2000) 525
  [arXiv:hep-ph/0002072].
  %%CITATION = HEP-PH 0002072;%%

%\cite{Khoze:2001xm}
\bibitem{Khoze:2001xm}
V.~A.~Khoze, A.~D.~Martin and M.~G.~Ryskin,
%``Prospects for new physics observations in diffractive processes at the  LHC
%and Tevatron,''
Eur.\ Phys.\ J.\ C {\bf 23} (2002) 311
[arXiv:hep-ph/0111078].
%%CITATION = HEP-PH 0111078;%%

%\cite{KRS}
\bibitem{KRS}
V.A.~Khoze, M.G.~Ryskin and W.J.~Stirling,
  arXiv:hep-ph/0504131.



%\cite{Sjostrand:2000wi}
\bibitem{Sjostrand:2000wi}
T.~Sj\"{o}strand, P.~Eden, C.~Friberg, L.~Lonnblad, G.~Miu, S.~Mrenna and E.~Norrbin,
%``High-energy-physics event generation with PYTHIA 6.1,''
Comput.\ Phys.\ Commun.\  {\bf 135} (2001) 238
[arXiv:hep-ph/0010017].
%%CITATION = HEP-PH 0010017;%%

\bibitem{cmstdr}
CMS Collaboration: The Trigger and Data Acquisition project, Volume II, Data Acquisition and High Level
Trigger, Technical Design Report, CERN/LHCC 2002-036.

\bibitem{atlastdr}
``ATLAS detector and physics performance. Technical design report. Vol. 2", CERN-LHCC-99-15. 

\bibitem{FP420}
M.G.Albrow et al., CERN-LHCC-2005-025 / LHCC - I - 015

%\cite{Butterworth:2002xg}
\bibitem{Butterworth:2002xg}
J.~M.~Butterworth, J.~P.~Couchman, B.~E.~Cox and B.~M.~Waugh,
%``KtJet: A C++ implementation of the K(T) clustering algorithm,''
Comput.\ Phys.\ Commun.\  {\bf 153} (2003) 85
[arXiv:hep-ph/0210022].
%%CITATION = HEP-PH 0210022;%%

%\cite{Butterworth:2002tt}
\bibitem{Butterworth:2002tt}
  J.~M.~Butterworth, B.~E.~Cox and J.~R.~Forshaw,
  %``W W scattering at the LHC,''
  Phys.\ Rev.\ D {\bf 65} (2002) 096014
  [arXiv:hep-ph/0201098].
  %%CITATION = HEP-PH 0201098;%%

\bibitem{famos}
J.~Kalliopuska {\it et al.}, HIP-2003-11/EXP

\bibitem{madx}
 F.~C.~Icelin {\it et al.}, MAD 9 Version 9, CERN-SL-2000-026 AP (2000).

%\cite{Pukhov:1999gg}
\bibitem{Pukhov:1999gg}
  A.~Pukhov {\it et al.},
  %``CompHEP: A package for evaluation of Feynman diagrams and integration  over
  %multi-particle phase space. User's manual for version 33,''
  arXiv:hep-ph/9908288.
  %%CITATION = HEP-PH 9908288;%%
%\cite{Pukhov:2004ca}
\bibitem{Pukhov:2004ca}
  A.~Pukhov,
  %``CalcHEP 3.2: MSSM, structure functions, event generation, batchs, and
  %generation of matrix elements for other packages,''
  arXiv:hep-ph/0412191.
  %%CITATION = HEP-PH 0412191;%%

\bibitem{KMRphot}
V.~A.~Khoze, A.~D.~Martin and M.~G.~Ryskin,
Eur.\ Phys.\ J.\ C {\bf 24} (2002) 459
[arXiv:hep-ph/0201301].

%\cite{Duhrssen:2005bz}
\bibitem{Duhrssen:2005bz}
  M.~Duhrssen, K.~Jakobs, P.~Marquard and J.~J.~van der Bij,
  %``The process g g $\to$ W W as a background to the Higgs signal at the LHC,''
  arXiv:hep-ph/0504006.
  %%CITATION = HEP-PH 0504006;%%

%\cite{Kleiss:1985yh}
\bibitem{Kleiss:1985yh}
  R.~Kleiss and W.~J.~Stirling,
  %``Spinor Techniques For Calculating P Anti-P $\to$ W+- / Z0 + Jets,''
  Nucl.\ Phys.\ B {\bf 262} (1985) 235.
  %%CITATION = NUPHA,B262,235;%%


\end{thebibliography}
\end{document}